\documentclass[twocolumn,preprintnumbers,superscriptaddress,amsmath,amssymb]{revtex4-1}

\usepackage[colorlinks=true,bookmarks=false,citecolor=blue,urlcolor=blue]{hyperref} 

\usepackage{amsmath,amssymb}
\usepackage{graphicx}
\usepackage{verbatim}
\usepackage{color}
\usepackage[english]{babel}
\usepackage{bm}
\usepackage{natbib}
\usepackage[symbol]{footmisc}
\usepackage{textcomp}

\usepackage{epstopdf}
\usepackage{setspace}

\usepackage{graphicx}% Include figure files
\usepackage{dcolumn}% Align table columns on decimal point
\usepackage{bm}% bold math
\begin{document}

\title{Broadband antireflection with halide-perovskite metasurfaces}

\author{Kseniia Baryshnikova}
\thanks{Corresponding Author}
\email{k.baryshnikova@metalab.ifmo.ru}
\affiliation{ITMO University, St.~Petersburg 197101, Russia}

\author{Dmitry Gets}
\affiliation{ITMO University, St.~Petersburg 197101, Russia}

\author{Tatiana Liashenko}
\affiliation{ITMO University, St.~Petersburg 197101, Russia}

\author{Anatoly Pushkarev}
\affiliation{ITMO University, St.~Petersburg 197101, Russia}

\author{Ivan Mukhin}
\affiliation{ITMO University, St.~Petersburg 197101, Russia}
\affiliation{Alferov University (former St. Petersburg Academic University),
St.~Petersburg 194021, Russia}

\author{Yuri Kivshar}
\affiliation{ITMO University, St.~Petersburg 197101, Russia}
\affiliation{Nonlinear Physics Center, Australian National University, Canberra ACT 2601, Australia}

\author{Sergey Makarov}
\thanks{Corresponding Author}
\email{s.makarov@metalab.ifmo.ru}
\affiliation{ITMO University, St.~Petersburg 197101, Russia}

\begin{abstract}
   Meta-optics based on optically-resonant dielectric nanostructures is a rapidly developing research field with many potential applications. Halide perovskite metasurfaces emerged recently as a novel platform for meta-optics, and they offer unique opportunities for control of light in optoelectronic devices. Here we employ the generalized Kerker conditions to overlap electric and magnetic Mie resonances in each meta-atom of MAPbBr$_3$ perovskite metasurface and demonstrate broadband suppression of reflection down to 4\%. We reveal also 
  that metasurface nanostructuring is also beneficial for the enhancement of photoluminescence. Our results may be useful for applications of nanostructured halide perovskites in photovoltaics and semi-transparent multifunctional metadevices where reflection reduction is important for their high efficiency.      
\end{abstract}

\maketitle

\section{Introduction}

Recently, meta-optics based on halide perovskites nanostructures~\cite{berestennikov2019active} has attracted a lot of attention
due to its numerous applications in structural coloring~\cite{gholipour17, makarov2017multifold}, enhanced luminescence~\cite{tiguntseva2018light}, tunable meta-pixels~\cite{gao2018lead, zhang2019lead}, optical encoding~\cite{fan2019resonance,zhizhchenko2020light}, lasing~\cite{tiguntseva2020room,huang2020ultrafast}, and photovoltaics~\cite{furasova2018resonant}. Various applications  of optoelectronic devises based on perovskites require suppression of reflection to achieve higher transparency in the sub-bandgap spectral range, but parasitic reflection limits the efficiency of photovoltaics devices, smart windows, and active glasses.
To reduce reflection, meta-optics employs the so-called Kerker effect and Huygens' metasurfaces~\cite{pfeiffer2013metamaterial,decker2015high,kuznetsov2016optically,staude2017metamaterial,de2018silicon}, that allow to suppress backscattering from each meta-atom. Such high-transmission metasurface designs were proposed and realized with standard semiconductor metasurfaces for  enhancing transparency of thin films in a broadband frequency range~\cite{decker2015high} and increasing absorption efficiency in active layers of photovoltaic devices~\cite{baryshnikova2016plasmonic}, as well as for bending light~\cite{yu2015high} and creating high-quality holographic images~\cite{wang2016grayscale}.

In this paper, we propose to employ the Kerker conditions for halide perovskite metasurfaces to achieve broadband  suppression of reflection with simultaneous enhancement of photoluminescence properties. We verify our theoretical concept 
experimentally by nanostructuring MAPbBr$_3$ perovskite thin films creating a periodical lattice of Mie-resonant perovskite nanoparticles. Optical characterization confirms that the optimized geometric parameters of the metasurface correspond to the reduction of reflectivity from 33\% down to 4\%, while photoluminescence intensity can be increased by at least 15\%.

\section{Results and Discussion}

%\begin{figure}
% \includegraphics[width=1.8\columnwidth] {Fig1.pdf}
% \caption{Fabrication of large-scale arrays of perovskite microlasers and their photoexcitation. Right panel: a close-up false-color SEM image (top) as well as a photograph of 1$\times$1 cm$^2$ array of perovskite microlasers.}
%  \label{thresholds}
%\end{figure}

\begin{figure*}[tb]
\centering
\includegraphics[width=1\linewidth]{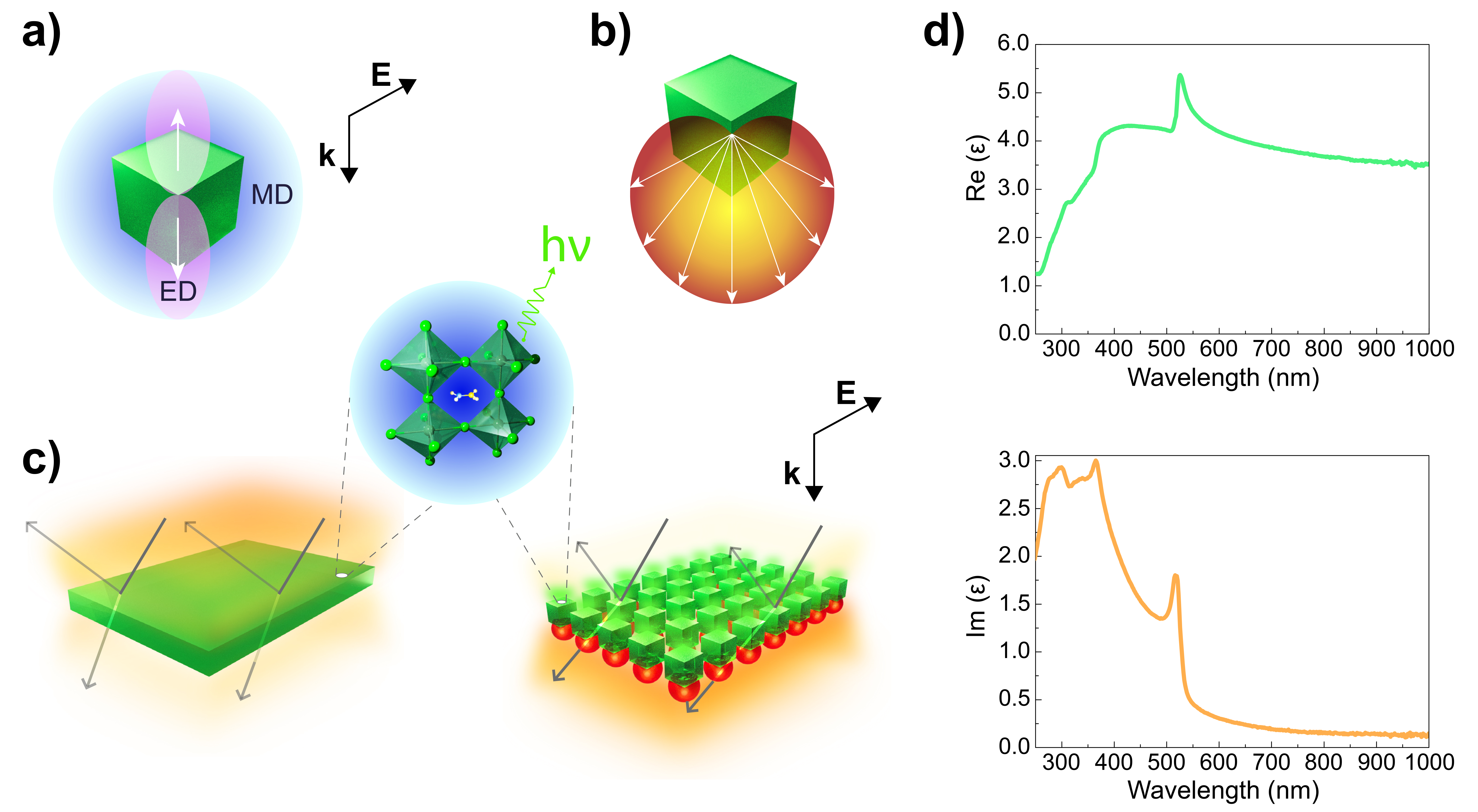}
\caption{\textbf{Concept of antireflected perovskite metasurfaces.} (a) Contribution of the electric and magnetic dipoles to scattering of a standalone nanoparticle. (b) Overall scattering pattern as a result of electric and magnetic dipoles interference. Energy is scattered mostly in the forward direction, making the particle to be analogue of a Huygens source. (c) Concept of Huygens metasurface based on perovskite particles. (d) Real and imaginary parts of dielectric permittivity of MAPbBr$_3$.}
\label{fig:figure1}
\end{figure*}

\textit{\textbf{Theoretical approach.}} As was shown previously, the halide perovskite nanoparticles (e.g. those made of MAPbBr$_3$) of various shapes can support excitation of pronounced Mie resonances in visible and near-infrared ranges~\cite{tiguntseva2018light, tiguntseva2018tunable, berestennikov2019beyond}. Thus, they should provide interesting physics based on interference between the excited Mie modes, including the Kerker effect.

The Kerker effect was originally predicted by Kerker et al.\cite{kerker1983electromagnetic} for spherical particles when dielectric permittivity equals to magnetic permeability. In this case electric dipole moment, excited in the particle, becomes equals in amplitude and in phase to magnetic dipole moment of this particle (so-called Kerker conditions) (see Figure~\ref{fig:figure1}a). Interference between electric and magnetic dipole radiation in this case is destructive in direction of backward scattering and constructive in direction of forward scattering. Thus, in the result, Kerker conditions cause to zero backward direction (see Figure~\ref{fig:figure1}b). While materials with electric permittivity equaled with magnetic permeability don't exist in nature, Kerker effect still can be observed for interaction of high-index nanostructures with light~\cite{gomez2011electric}. This effect is widely used in all-dielectric nanophotonics due to highly transparent features of metasurfaces based on such type of high-index nanostructures~\cite{kruk2020tailoring}. %Metasurface based on such nanoparticles is named usually as a Huygens metasurface, because each element in this metasurface acts as a Huygens source (see Figure~\ref{fig:figure1}c)~\cite{decker2015high}.
Frequently high refractive index is treated as a required condition for all-dielectric nanophotonics structures, because in case of low refractive index resonant response of particles is suppressed and interaction with light is rather weak. However, it was shown recently, that low contrast between materials of particle and its environment can cause to a broadband Kerker effect~\cite{dezert2017isotropic, terekhov2019broadband}. Interestingly, while in the references above contrast between materials was reduced artificially, halide perovskite has naturally low refractive index which can cause to a broadband Kerker effect. It means that metasurface based on perovskite meta-atoms can be non-reflected and almost transparent, besides of energy absorbed in the nanoparticles volume.
The non-zero absorbance of particle's material also influence on the scattering properties. In Ref.~\cite{yang2018nonradiating}, it was shown that increasing of absorbance can destroy directive scattering even when multipoles condition is satisfied.

To optimize metasurface geometry, at first we used approximation of spherical particles which allows us using of fully analytical Mie theory to describe optical response of perovskite nanoparticles~\cite{wriedt2012mie}. We consider standalone nanoparticle placed in homogeneous environment with refractive index equals to 1.51  which corresponds to refractive index of glass or polymer. It was shown that for particles with radii up to 100 nm electric dipole contribution to scattering cross section is predominant for all the visible range (Figure S1). When radius of nanoparticle is growing further, magnetic dipole contribution is increasing simultaneously with electric quadrupole and magnetic quadrupole contributions. Because of this fact, not original Kerker conditions, but generalized Kerker conditions, which includes both dipoles and quadrupoles contribution, should be considered~\cite{liu2018generalized, pors2015unidirectional}. Let's consider them in details.

\begin{figure*}[t!]
\centering
\includegraphics[width=1\linewidth]{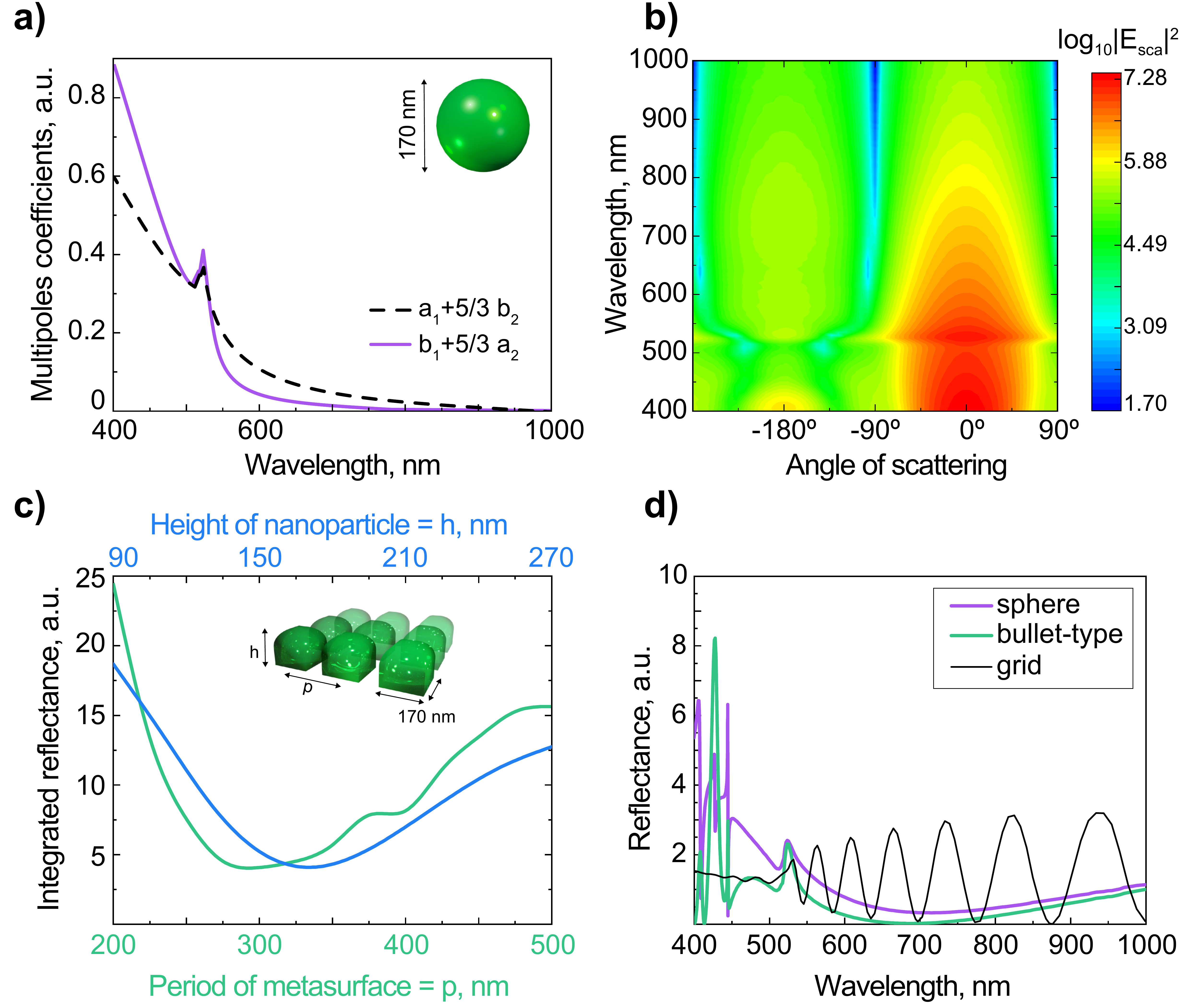} 
\caption {\textbf{Analytical and numerical modeling}. (a) Illustration of generalized Kerker conditions satisfaction for a sphere with diameter of 170 nm (see Eq.~7). (b) Logarithm of numerically calculated scattering amplitude for the perovskite nanocube placed into the homogeneous environment as a function of wavelength and observing angle. (c) Reflection from a perovskite metasurface on a glass substrate integrated in the spectral range $\lambda$=400-1000~nm as a function of perovskite parallelepiped nanoparticle height \textit{h}. (d) Reflection spectra for metasurface based on spherical (violet line) / bullet-type (green line) / needle-type particles (black line) placed on a glass substrate.}
\label{fig:theory}
\end{figure*}

The scattered far field of particle of arbitrary shape $E_{sca}$ is a spherical wave which can be described by following projections in spherical coordinates:
\begin{equation}
    E_{s\theta}=\frac{e^{ikr}}{-ikr}cos{\varphi}\cdot S_2(cos \theta)
\end{equation}
\begin{equation}
    E_{s\varphi}=\frac{e^{ikr}}{ikr}sin{\varphi}\cdot S_1(cos \theta)
\end{equation}
with the scattering amplitudes $S_1$ and $S_2$:
\begin{equation}
    S_1(cos \theta)= \sum_{n=1}^{\infty} \frac{2n+1}{n(n+1)}(a_n\pi_n+b_n\tau_n)
\end{equation}
\begin{equation}
    S_2(cos \theta)= \sum_{n=1}^{\infty} \frac{2n+1}{n(n+1)}(a_n\tau_n+b_n\pi_n)
\end{equation}
Here $a_n$ ($b_n$) is scattering coefficient of Mie theory, which defines  contribution on electric (magnetic) multipole on n-th order, and $\tau_n$ and $\pi_n$ are functions which are related with Legendre functions of the first order:
\begin{equation}
    \tau_n=\frac{dP_n^1(cos \theta)}{d\theta}
\end{equation}
\begin{equation}
    \pi_n=\frac{P_n^1(cos \theta)}{sin \theta}
\end{equation}
To obtain suppression of backward scattering, $S_1$ and $S_2$ should be zero when $\theta=180^{\circ}$. Taking into account first two Mie coefficients which correspond to dipoles and quadrupoles contributions, and solving the simple system of linear equations, one can express generalized Kerker condition as follows:
\begin{equation}
    a_1+\frac{5}{3}b_2=b_1+\frac{5}{3}a_2
\end{equation}

Simple optimization using experimental data for perovskite dielectric function shows that diameter of the sphere equaled to 170 nm corresponds to satisfying of generalized Kerker coтditions in the vicinity of nanoparticle resonance and it is expected to satisfy them approximately in all the visible range (see Figure~\ref{fig:theory}a).

\begin{figure}[b!]
\centering
\includegraphics[width=1\linewidth]{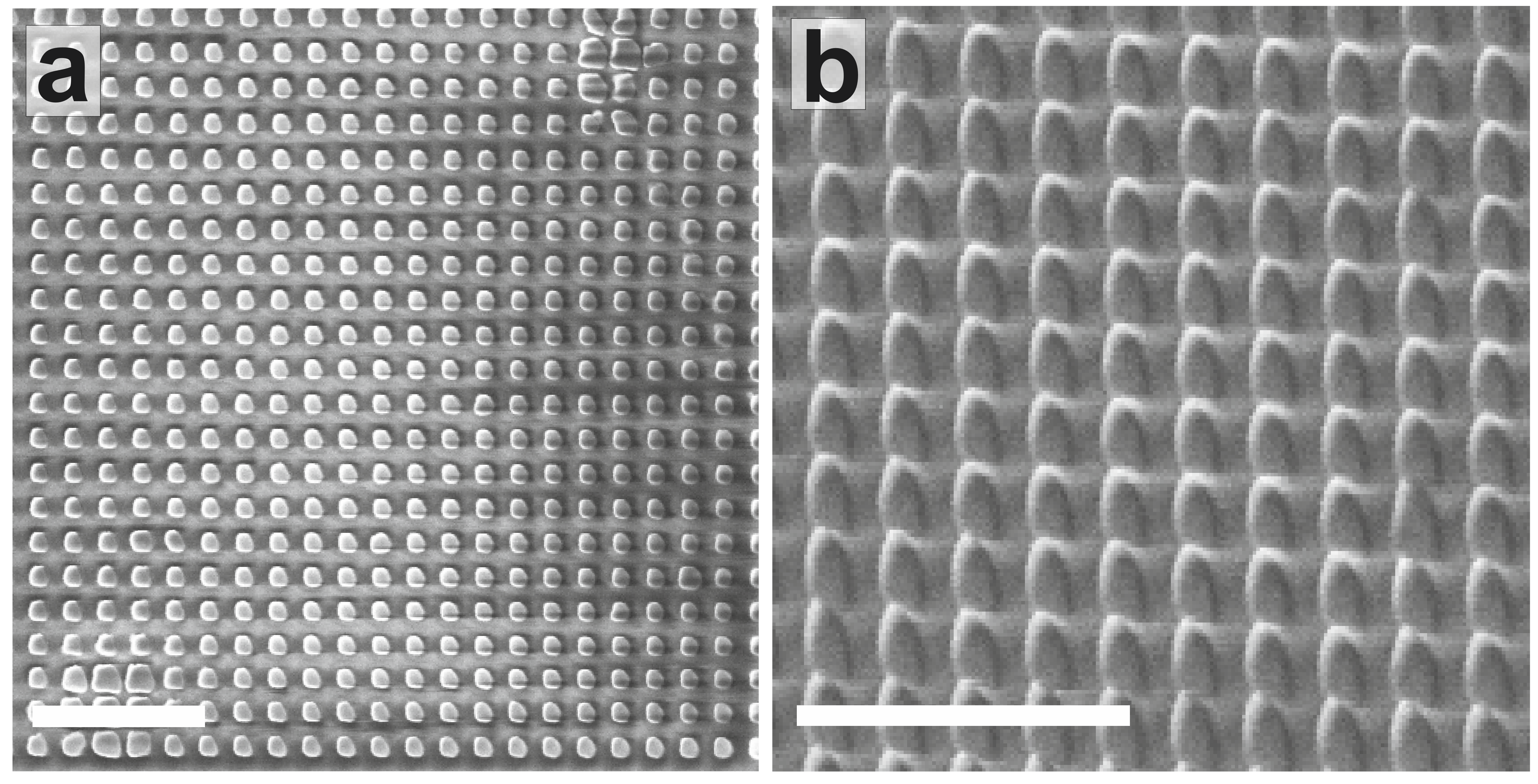}
\caption{\textbf{Example of fabricated perovskite metasurface.} SEM images of MAPbBr$_3$ perovskite metasurface from top (a) and 30$^o$ side (b) views. Scale bars are 2~$\mu$m.}
\label{fig:SEM}
\end{figure}

\begin{figure*}[t]
\centering
\includegraphics[width=0.99\linewidth]{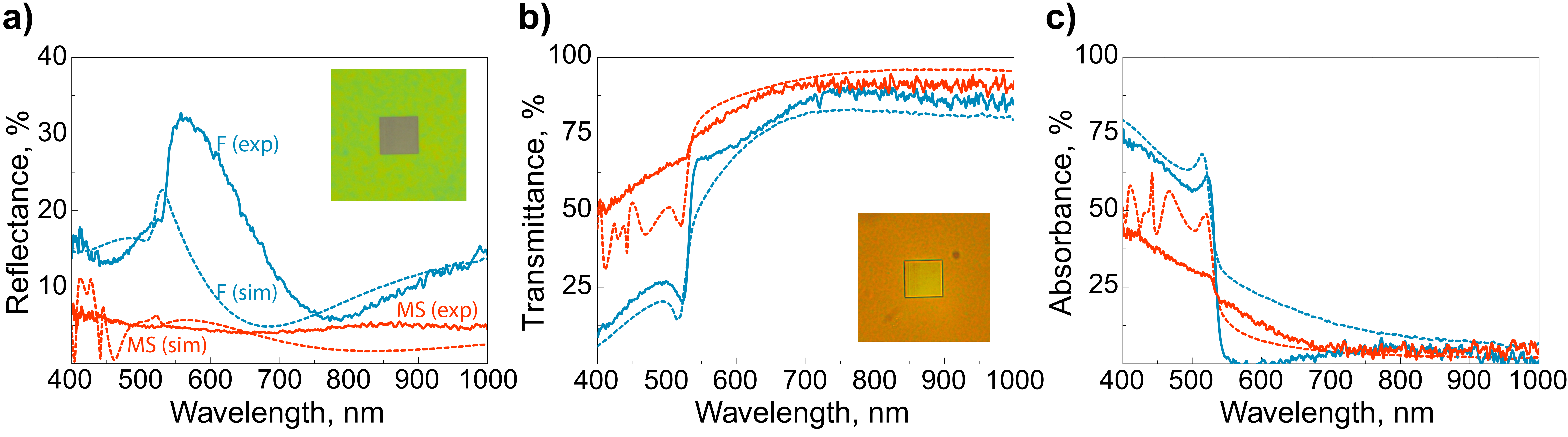}
\caption{\textbf{Optical characterization.} Experimentally measured spectra of reflectance (a) and transmittance (b) for 170-nm perovskite film (F, blue curve) and metasurface (MS, red curve). Insets show optical images in reflection and transmission, respectively. (c) Photoluminescence spectra from the MAPbBr$_3$ metasurface and thin film.}
\label{fig:Exp1}
\end{figure*}

{\bf Numerical simulations}. While results for sphere are encouraging, this shape of particle is hardly achievable for manufacturing. Based on used manufacturing methods, bullet-type nanoparticle could be used for experiment (see inset in Fig. \ref{fig:theory}c). We predicted that scattering optical properties of bullet-type nanoparticle are similar to properties of spherical nanoparticle due to similar multipoles enhancement. 

To verify this prediction, we simulated numerically the scattering pattern of the bullet-type nanoparticle placed in the homogeneous environment with refractive index equaled to 1.51. It corresponds to case where particle is placed on the glass substrate and covered by the polymer and immersion oil with similar optical density. Figure~\ref{fig:theory}b represents scattering amplitude for considered nanoparticle in specific angle in a wide wavelength range from 400 to 1000 nm. It is clearly seen that in the whole wavelength range energy is scattered predominantly in ''zero'' angle which correspond to the forward direction, so it proves our original concept. Also it should be noted that for wavelength smaller than 520 nm scattering in the backward direction increases. It corresponds to the case of absorptive nanoparticle mentioned above. Besides of this wavelength range, backscattering is close to zero so it is reasonable to expect that metasurface based on such nanoparticles possesses small reflectance.

To prove it, we simulated interaction of metasurface based on bullet-type nanoparticles placed on the top of glass substrate with normally incident linearly polarized light and sweep the geometrical parameters of metasurfaces. While base size of particle stayed the same, height of nanoparticle and period of metasurface can vary. Results of simulations, gathered in Figure \ref{fig:theory}c show that height of nanoparticle (blue line) equaled to 170 nm and period of metasurface (green line) equaled to 300 nm correspond to the minimum of integrated reflectance.

However, this parameters of metasurface could correspond to the just optimal filling factor which could be easily satisfied in the metasurfaces based on the particles of other shape. To show that considered broadband anti-reflectance is not an effect of the special nanoparticles shape or suitable filling factor, we compared results of metasurfaces of different shapes but the same filling factor in Figure \ref{fig:theory}d. Spherical nanoparticles don't provide the gradient of refractive index between the air and the substrate, but have almost the same properties with the bullet-type nanoparticle due to similar multipoles excitation (see Figure S2). And vice versa, metasurface consisted of long needles with the same filling factor and nanoparticle volume (the height of each needle is 2720 nm, the base side is 42.5 nm, the period is 75 nm; ''grid'' line in Figure \ref{fig:theory}d) corresponds to the strong oscillations on reflectance which cause to the increasing of the integral reflectance. 

To take into account all the parameters of system in an experiment, additional optimization was performed, where a width of a covering polymer layer was added as a parameter of the optimization. This layer can work as additional anti-reflective coating and could increase antireflective properties, too. Performing careful optimization using CST Studio suit software, following geometry were proposed for fabrication: bullet-type nanoparticles with base side of 170 nm are placed in the nodes of the square lattice with period 300 nm of on the glass substrate and covered by the polymer of width 270 nm (i.e. 100 nm above the upper level of cubes). This configuration corresponds to minimal integral reflection $\int R(\lambda) d\lambda$ in the wavelength range 400-1000 nm. Here $\lambda$ is a wavelength, $R(\lambda)$ is reflection coefficient at the $\lambda$. Figure \ref{fig:Exp1}c shows simulated reflection spectra for bare perovskite layer and for metasurface. It is worth to emphasize that while reflectance of metasurface is much smaller than reflectance of bare layer in all the visible range, absorption is not reduced dramatically at wavelengths above 520 nm. It is important for effective luminescence of resulted perovskite metasurface.

\textit{\textbf{Experimental results.}} Basing on the developed design of a broadband antireflective metasurface, we realize it experimentally by employing focused ion beam (FIB) lithography for milling MAPbBr$_3$ thin film of thickness 170~nm. The details of the film synthesis and deposition are given in Methods. Scanning electron microscopy (SEM) images of the fabricated metasurface are shown in Fig.~\ref{fig:SEM}a,b, revealing high quality of the design and desired period 400~nm.

According to the optical images in reflection (Fig.~\ref{fig:Exp1}a) and transmission (Fig.~\ref{fig:Exp1}b) modes, the fabricated metasurface exhibits pronounced anti-reflective properties. Namely, spectroscopic measurements from the samples confirm the theoretical predictions and reveal reduction of reflectance from 33\% down to 4\%. The achieved regime antireflection is extremely broadband keeping the 4--6\% level over whole visible and part of near-infrared ranges. Remarkably, the slightly increased absorption or scattering in range $\lambda$=550--700~nm (see Fig.\ref{fig:Exp1}c) in the nanostructured perovskite film do not strongly affect the reflectance and transmittance. In opposite, the range of $\lambda$=400--550~nm corresponds to strong light absorption in MAPbBr$_3$ perovskite, resulting in big changes after the nanostructuring owing to simple reduction of the lossy material area by more than 4 times.

On one hand, the removal considerable part of the film should result in immediate drop of photoluminescence (PL) integral power. On the other hand, according to the simulations, the emission range ($\lambda\approx$~520--560~nm) overlaps with ED, MD, and EQ Mie resonances in the nanoparticles (see Fig.S1,S2). Generally, the coupling of emission in perovskite nanoparticles with such resonances results in the PL enhancement as shown in previous studies~\cite{tiguntseva2018light, berestennikov2019active} because of Purcell effect.
However, preliminary tests of PL from the metasurfaces upon irradiation by UV part ($\lambda$=365~nm) of a mercury lamp spectrum reveal strong quenching of PL signal. Indeed, because FIB lithography implies material processing by Ga$^+$-ions, which results in defects generation in the perovskite, according to the previous studies~\cite{wang2019perovskite}. 

\begin{figure}[t]
\centering
\includegraphics[width=1\linewidth]{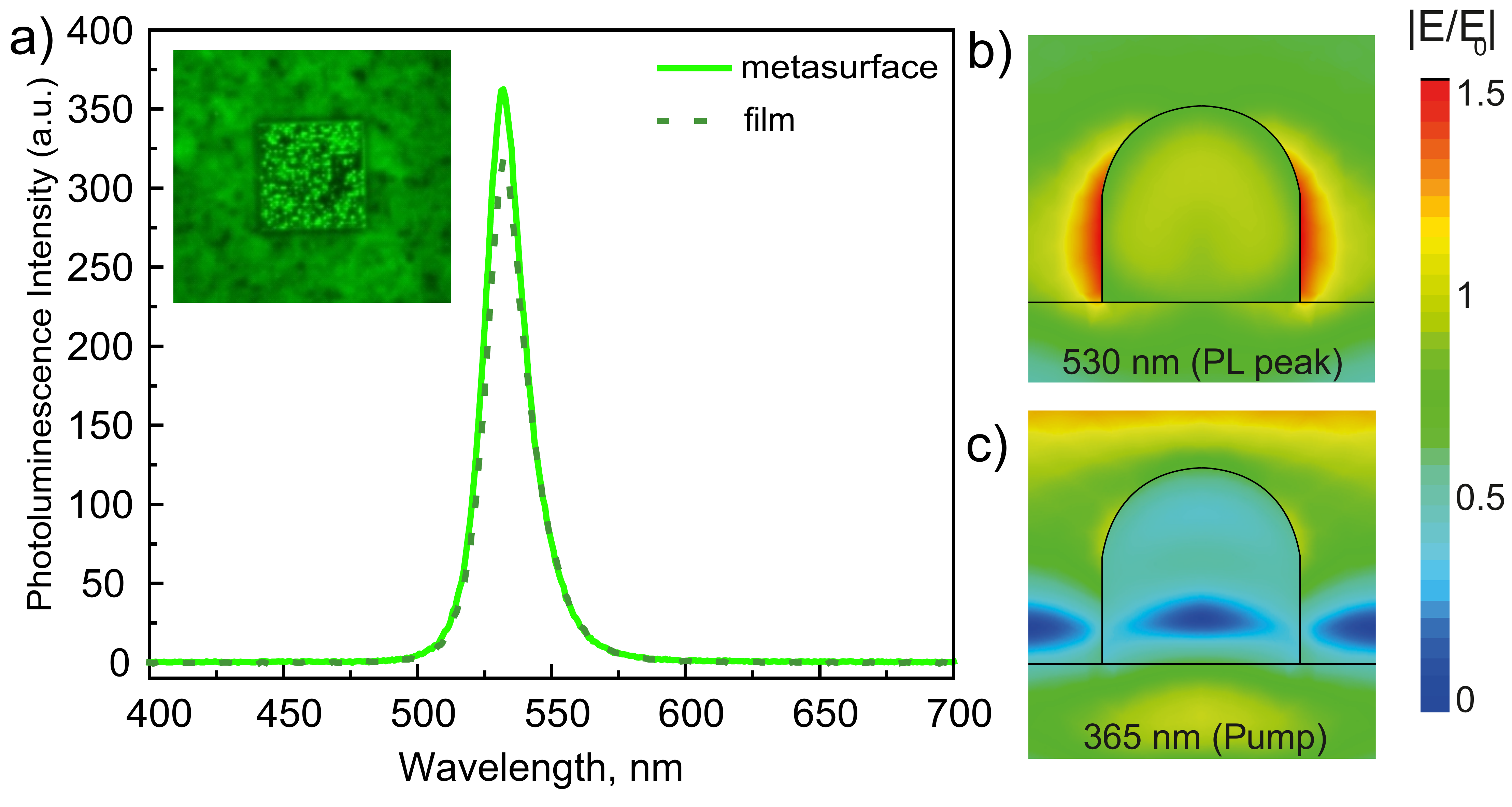}
\caption{\textbf{Photoluminescence enhancement.} (a) Measured spectra of photoluminescence for 170-nm perovskite film (dash curve) and metasurface (solid curve). Inset shows photoluminescent image from the MAPbBr$_3$ metasurface. Numerically calculated electric field distribution in a unit cell of the metasurface at emission wavelength 530~nm (b) and pump wavelength 365~nm (c).}
\label{fig:Exp3}
\end{figure}

In order to overcome this problem, we propose the defects passivation by perovskite exposition in vapors of isopropyl alcohol (IPA) outside the glove box. In our experiments, the sample with metasurface is placed in a small Petri dish which is put in the bigger Petri dish and heated on the hot plate at 60~$^{\circ}$C for 2 minutes. Then, 50~$\mu$l of the dissolved solution of MABr in IPA with concentration 20~g~mL$^{-1}$ (V~=~0.5~ml) was dripped onto the bottom of the bigger Petri dish close to the small one, after that covered with the preheated lid of the big Petri dish and kept in the mist at 60~$^{\circ}$C for 10 minutes. 
After that, we observe recovery of the PL intensity. Moreover, the metasurface demonstrates even 15\% higher PL signal compared to the neighbour smooth area of the film (Fig.~\ref{fig:Exp3}a). However, taking into account 3 times less volume of emitting material as compared with the smooth film around, we can conclude that we improve PL yield per volume by more than 3 times. 

According to the previous studies, it is related to the Purcell effect in the resonant perovskite nanoparticles~\cite{tiguntseva2018light, berestennikov2019active}, where ED, MD, and EQ are spectrally overlapped around the emission wavelength. Indeed, our simulations of near-field distribution around nanoparticle in each unit cell of the metasurface reveal up to 1.5-fold electric field enhancement (or up to 2.25 times of squared field enhancement) at the emission wavelength (see Fig.~\ref{fig:Exp3}b), which is an indication of enhanced Purcell factor according to the reciprocity theorem~\cite{krasnok2016demonstration}. Regarding the near-field distribution around the nanoparticle at pump wavelength, we did not observe any significant enhancement (Fig.\ref{fig:Exp3}).

%\section{Conclusion}

In summary, we have demonstrated that nanostructuring of halide perovskite thin films results in a strong suppression of reflection as well as enhancement of photoluminescent properties. In particular, we have fabricated perovskite metasurfaces with reflection less than 4\% and transparency around 90\% in a broad spectral range, with photoluminescence signal increased by 15\% relative to an unpatterned film of the same thickness. Our design is based on the concept of the generalized Kerker effect realized in individual perovskite nanocuboids forming a periodic lattice. Additional chemical post-processing has allowed for passivating defects in the perovskite film, and, thus, eliminating the harmful effect of PL quenching after the FIB nanostructuring. 

We believe that our approach can be applied to a broad range of perovskite-based light-emitting structures with the refractive index close to that of MAPbBr$_3$. Thus, a combination of novel strategies for geometrical and chemical optimisations of perovskites paves the way for multifunctional perovskite-based optoelectronic devices with high efficiency and high transparency. More specifically, this concept can be applied for improving light-emitting solar cells~\cite{kim2017peroptronic, gets2019light}, when the nanostructured perovskite layer operates as an active semi-transparent layer as well as an additional layer for a solar cell to convert the incident UV light into green photons for improving the device efficiency. Finally, the proposed design of anti-reflecting perovskite metasurfaces can be up-scaled by both mechanical nanoimprint~\cite{wang2016nanoimprinted, makarov2017multifold} and laser printing~\cite{zhizhchenko2019single, zhizhchenko2020light} technologies.

\section{Methods}\mbox{}

\textbf{Materials.} Lead(II) bromide (PbBr$_2$,  99.998$\%$ trace metals basis, Puratronic, Alfa Aesar), methylammonium bromide (MABr, 99.5$\%$ DYESOL), dymethyl sulfoxide (DMSO, anhydrous 99.5$\%$) and diethyl ether (anhydrous, Sigma-Aldrich) were used as received without additional purification.

\textbf{Preparation of perovskite precursor solutions.} \textit{MAPbBr$_3$ solution.} PbBr$_2$ (293.6 mg) and MABr (89.57 mg) were subsequently dissolved in 1 ml of anhydrous DMSO. All the solutions were prepared in a N$_2$-filled glove box operating at 0.1 ppm of both O$_2$ and H$_2$O. 

\textbf{Thin films deposition.}
 The glass substrates (18x18 mm$^{2}$) were subsequently ultrasonicated in acetone and 2-propanol for 5 min, respectively, then washed with deionized water, and dried with dry air and finally cleaned with O$_3$ for 10 min. Perovskite films were spin-coated onto glass substrates by anti-solvent dripping method in the glove box. At first the galss substrate was fixed on the chuck of the sprincoater. Then 100 $\mu$l of perovskite ink was dripped into the center of the substrate. The spincoating cycle had following steps: at first step spincoater reached rotational speed of 1000rpm for 5 s at kept that speed for 40 s then it accelerates up to 3000 rpm for 3 s and kept this speed for 40s. At 32th s of the second cycle, 1 ml of diethyl ether was dripped onto the substrate. The films were annealed on the hot plate at 50 $^{\circ}$C for 5 min and then at 100 $^{\circ}$C for 10 min.

\textbf{Metasurface fabrication}
Metasurfaces were milled from the thin perovskite film with the use of focused ion Ga$^+$ beam. Structures fabrication was carried out with the system of crossed electron and ion beams implemented in Neon 40 (Carl Zeiss). For the milling process 10~pA ion current was used, which provides material etching without re-deposition~\cite{evstrapov2012ion} The result of etching was visualized via scanning with electron beam.

\textbf{Optical measurements}
Measurements of reflection (R) for thin film and metasurfaces was carried out in optical microscope (Carz Zeiss) with an objective $\times$100 NA = 0.9 focused and collected white light from a halogen lamp. Transmittance was measured on the same setup but with illumination of the samples from bottom by a condenser lens. The signal spectra were collected by a optical fiber with 600 $\mu$m core delivering light to a spectrometer (Ocean Optics, QE Pro).

\textbf{Analytical results and numerical simulations.} 
Analytical predictions were made by employing the original Mie calculator implemented in Matlab. Numerical properties of standalone nanoparticles were performed via Comsol Multiphysics. Numerical simulations of metasurface optical properties were performed via CST Microwave studio due to its high performance in tasks of periodical structures simulations.

\section{Acknowledgements}

This work was supported by the Russian Science Foundation (project no 19-73-30023), the Australian Research Council (grant DP200101168), and the Strategic Fund of the Australian National University.

\bibliography{refs}

\end{document}